\documentclass[prb,twocolumn,superscriptaddress,showpacs,amsmath,amssymb]{revtex4}
\usepackage{graphicx}
\usepackage{latexsym}
\usepackage{amsmath}
\usepackage{amssymb}
\usepackage{amsfonts}
\usepackage{color}

\usepackage{amssymb}
\usepackage{bm}
\usepackage{verbatim}
\usepackage[cp1251]{inputenc}

\usepackage[colorlinks,bookmarks=false,citecolor=blue,linkcolor=red,urlcolor=blue]{hyperref}
\usepackage{color}
\usepackage[all]{hypcap} % let hyperlinks correctly point to figures rather than their captions;

\newcommand{\ds}{\displaystyle}
\begin{document}
% You should use BibTeX and apsrev.bst for references
%\bibliographystyle{apsrev}
%\preprint{}

\title{Long--range triplet proximity effect in multiply connected ferromagnet-superconductor hybrids}

  \author{A.V. Samokhvalov}
  \affiliation{Institute for Physics of Microstructures, Russian Academy of Sciences, 603950 Nizhny Novgorod, GSP-105, Russia}
  \affiliation{Lobachevsky State University of Nizhny Novgorod, Nizhny Novgorod 603950, Russia}
  \author{J.W.A. Robinson}
  \affiliation{Department of Materials Science \& Metallurgy, University of Cambridge, CB3 0FS Cambridge,United Kingdom}
  \author{A.I. Buzdin}
  \affiliation{Department of Materials Science \& Metallurgy, University of Cambridge, CB3 0FS Cambridge,United Kingdom}
  \affiliation{University Bordeaux, LOMA UMR-CNRS 5798, F-33405 Talence Cedex, France}
  \affiliation{Sechenov First Moscow State Medical University, Moscow, 119991, Russia}

%\date{\today}

\begin{abstract}
Applying the linearized Usadel equations, we consider the
nucleation of superconductivity in multiply connected mesoscopic
superconductor/ferromagnet (S/F) hybrids such as a thin
superconducting ring on a ferromagnet with a uniform in-plane
magnetization $M$ and a spin-active S/F interface. We demonstrate
that the exchange field in F provokes a switching between
superconducting states with different vorticities which may
increase the critical temperature ( $T_c$ ) of the superconductor
in a magnetic field. We study the interplay between oscillations
in $T_c$ due to the Little--Parks effect and oscillations in $T_c$
induced by the exchange field. Furthermore, we analyse the
influence of long-range spin-triplet correlations on the switching
between different vorticities.
\end{abstract}
 % Mesoscopic and nanoscale systems
%%74.78.-w  % Superconducting films and low-dimensional structures
%% insert suggested keywords - APS authors don't need to do this
%%\keywords{}

\maketitle

\section{Introduction}

%Recently, research activity in the physics of hybrid structures
%containing singlet superconducting (S) and ferromagnetic (F)
%materials (see reviews
%\cite{Buzdin-RMP05,Izyumov-UFN02,Bergeret-RMP05,Linder-NatPhys15,Eschrig-RPP15}
%for details) provided an impetus for the studies of the
%long-range proximity effects in the ferromagnets \cite{Giroud-PRB98,Sosnin-Petrashov-PRL06,Keizer-Nature06,%
%Wang-NatPhys10-Co_wire,Robinson-Sci10,Robinson-Buzdin-PRL10,Khaire-Birge-PRL10,Klose-PRL12}
%related to the odd-frequency superconductivity
%\cite{Berezinskii-JETPL74} and generation of spin-polarized
%triplet Cooper pairs near SF interfaces
%\cite{Bergeret-PRL01,Kadigrobov-Shekhter-EL01}.

In hybrid systems containing superconducting and ferromagnetic
metals (see reviews
\cite{Buzdin-RMP05,Izyumov-UFN02,Bergeret-RMP05,Blamire-Robinson-CM14,Linder-NatPhys15,Eschrig-RPP15}),
long-range correlations are induced in the ferromagnets \cite{Giroud-PRB98,Sosnin-Petrashov-PRL06,Keizer-Nature06,%
Wang-NatPhys10-Co_wire,Robinson-Sci10,Robinson-Buzdin-PRL10,Khaire-Birge-PRL10,Klose-PRL12}
related to odd-frequency superconductivity
\cite{Berezinskii-JETPL74} and spin-polarized triplet Cooper pairs
\cite{Bergeret-PRL01,Kadigrobov-Shekhter-EL01}. Such long-range
spin-triplet pairs are not destroyed by a ferromagnetic  exchange
field and can penetrate a ferromagnet over long distances
exceeding a singlet pair coherence length. The reduction of a Co
nanowire's resistance in contact with a superconductor observed in
\cite{Giroud-PRB98,Kompaniiets-APL14} as the temperature is
decreased below the superconducting transition, demonstrated
evidence for a long--range proximity effect. Evidence for electron
pair conversion from spin-singlet to spin-triplet has recently
been demonstrated via observations of long-ranged supercurrents in
S/F/S Josephson junctions with magnetically inhomogeneous S/F
interfaces
\cite{Robinson-Sci10,Robinson-Buzdin-PRL10,Witt-Robinson-PRB12,Robinson-SciRep12,Robinson-EPL14-HM,Robinson-PRB14-Cr,Robinson-NatComm14,Khaire-Birge-PRL10,Klose-PRL12},
transition temperature measurements of S/F1/F2 spin valves
\cite{Leksin-PRL12,Wang-PRB14,Singh-PRX15,Gu-Robinson-PRL15,Srivastava-PRAp17,Gu-Robinson-APLMat14},
density of states measurements on S/F systems
\cite{SanGiorgio-PRL08,Usman-PRB11,Boden-PRB11,Robinson-PRB12,Robinson-PRB14,Robinson-PRB15,Robinson-NatComm15,Robinson-PRB17}
and ferromagnetic resonace
\cite{Robinson-NatMat17,Jeon-Robinson-PRB19}. The optimal
condition for pair conversion in a dirty ferromagnet is realized
when the exchange field $h$ is inhomogeneous on scale of the
coherence length $\xi_f = \sqrt{\hbar D_f / h}$ in the F
\cite{Houzet-Buzdin-PRB07}, where $D_f$ is the electron diffusion
constant. Under appropriate conditions, odd-frequency spin-triplet
correlations manifest as an intrinsic paramagnetic Meissner state
\cite{Bergeret-PRB01,Yokoyama-Tanaka-PRL11_Meissner-effect-SN-SA,Mironov-PRL12_Meissner-effect}.
The existence of an anomalous Meissner response has been observed
via a depth-resolved measurements of the local magnetic fields in
Au/Ho/Nb using low-energy muons \cite{Bernardo-PRX15-paramagnetic
Meissner state}.

%
%%%%%%%%%%%%%%%%%%%%%%%%%%%%%%%%%%%%%%%%%%%%%
\begin{figure*}
\includegraphics[width=0.35\textwidth]{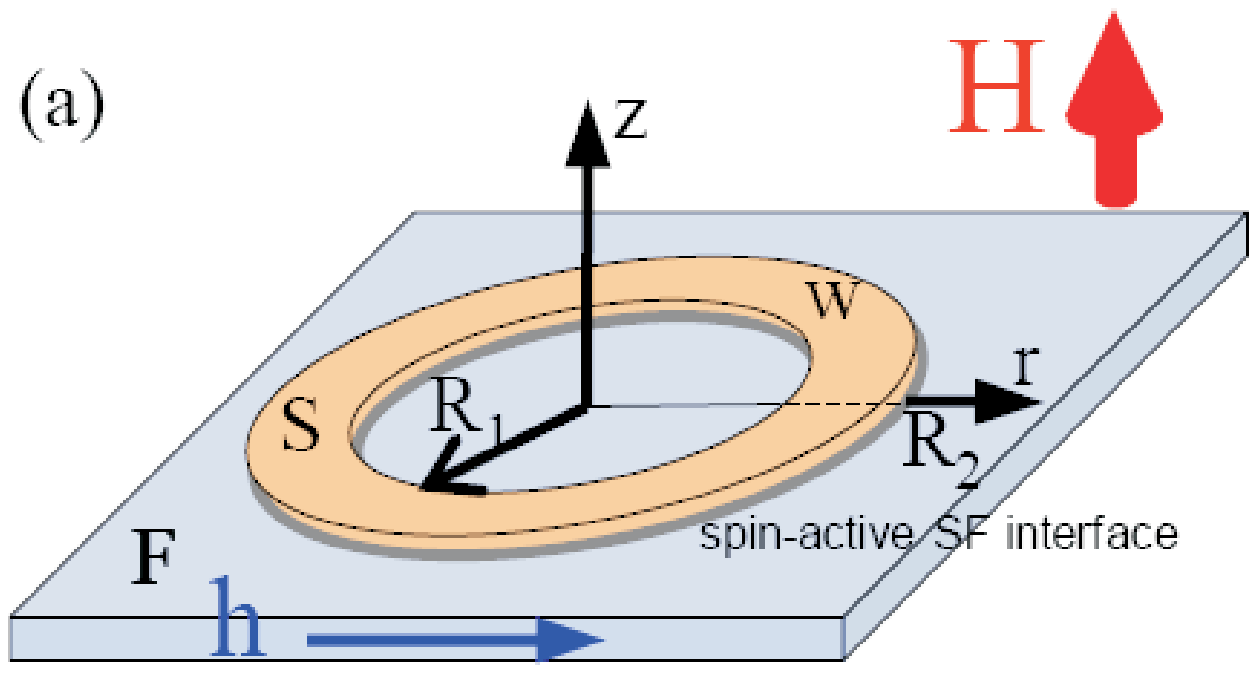}\hspace{1 cm}
\includegraphics[width=0.35\textwidth]{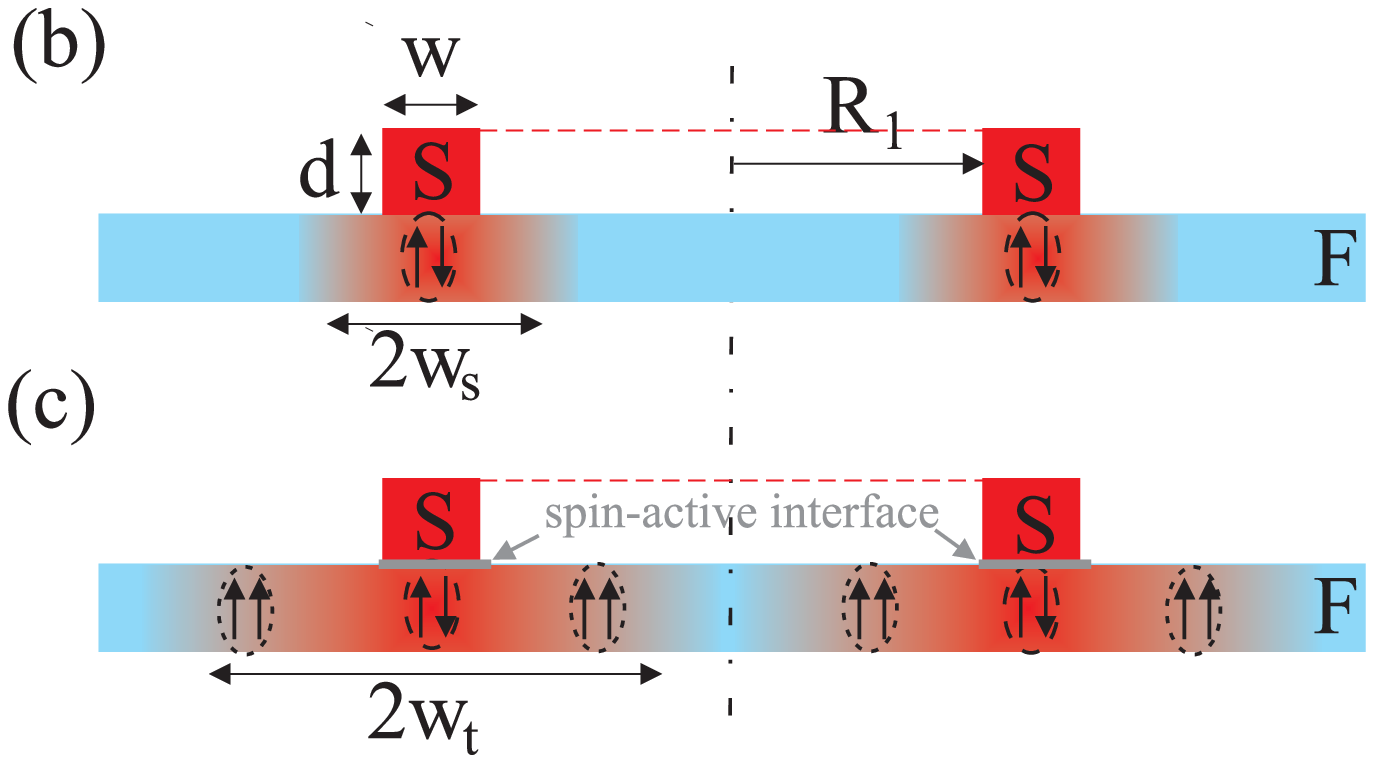}
\caption{(Color online) (a) A superconducting ring (S) on the
surface of a ferromagnetic plate (F) with a uniform exchange field
$\mathbf{h}$ and a spin-active S/F interface. Here $R_1$ ($R_2$)
and $w=R_2-R_1$ are the inner (outer) radius and the width of the
S ring, $(r,\theta,z)$ is the cylindrical coordinate system.
The external magnetic field $\mathbf{H}$ is applied along the $z$ axis.
(b) Short-range proximity--induced superconductivity in a F metal.
(c) Long-range proximity induced superconductivity in a F metal due
the precence of a spin-active S/F interface. Arrows indicate the
spin structure of pair correlations in F.} \label{Fig:1-Ring}
\end{figure*}
%%%%%%%%%%%%%%%%%%%%%%%%%%%%%%%%%%%%%
%

Long-range proximity effect have mainly focused on mesoscopic
ferromagnetic wires attached to a superconductor
%\cite{Giroud-PRB98,Sosnin-Petrashov-PRL06,Keizer-Nature06,Wang-NatPhys10-Co_wire}
or layered (diffusive) S/F structures.
%\cite{Robinson-Sci10,Robinson-Buzdin-PRL10,Khaire-Birge-PRL10,Klose-PRL12}.
Here we propose using multiply connected hybrid S/F structures
such as a thin narrow superconducting ring on a ferromagnetic
plate (see Fig.~\ref{Fig:1-Ring}(a)) to study
proximity--induced long--range triplet correlations. A
distinguishing feature of multiply connected S/F hybrids with the
proximity effect is the generation of a vortex state with a
nonzero orbital angular momentum $L$ (vorticity) even in the
absence of an external magnetic field \cite{Samokhvalov-PRB07v}.
The damped oscillations of the amplitude of singlet
superconducting correlations in a ferromagnet in the direction
perpendicular to the S/F interface
\cite{Buzdin-Bulaevskii-JETPL82,Buzdin-Kupriyanov-JETPL91,Buzdin-Kupriyanov-JETP92}
generates the additional phase modulation of superconducting order
parameter $\Delta \sim \exp(i L \theta)$ on azimuthal angle
$\theta$ and may induce spontaneous supercurrents. The angular
momentum $L$ of the pair wave function determines the vorticity of
the state. The interplay between orbital \cite{Ginzburg-ZETF56}
and the exchange \cite{SaintJames-SCII} effects may result in a
switching between the states with different vorticities
$L$, as the ring radius $R_1$ increases. Transitions between
states with different $L$ result in a nonmonotonic dependence of
the critical temperature $T_c$ on $R_1$ and $h$.

The Little-Parks (LP) effect (the periodic oscillations of
the critical temperature $T_c$ of a hollow superconducting
cylinder in an applied magnetic field $H$
\cite{Little-Parks-PRL62,Little-Parks-PR64}) is an
extremely sensitive experimental tool for studying interference
phenomena in multiply connected superconducting systems. The
orbital effect results in switching between states with different
vorticity $L$ and manifests itself in oscillations of the
phase--transition line $T_c(H)$. The interplay between $T_c$
oscillations due to LP effect and oscillations
due to $h$ in multiply connected S/F hybrids with a uniform
magnetization was shown to be accompanied by breaking of
the strict periodicity of $T_c(H)$ oscillations and shifts in
$T_c$ maximum to finite $H$ \cite{Samokhvalov-PRB09v}.
Similar results were obtained later in
\cite{Krunavakarn-PhysC13v,Krunavakarn-PhysC16v} for the case of a
spiral exchange field distribution in a ferromagnet. Note,
however, that in the case of a magnetic spiral, equal--spin
triplet pairs do not arise, and long-range proximity induced
superconductivity is absent \cite{Eschrig-RPP15,Linder-PRB10}.

The aim of our paper is to examine the long-range proximity
effect on switching between vortex states in multiply connected
S/F hybrids. We expect that the superconducting ground state in
such a geometry should be strongly influenced by the presence of
equal--spin triplet pairs in F. We focus on the behavior of $T_c$
for superconducting states with different vorticities, and
study the influence of the equal--spin triplet components on LP
oscillations. We start from a qualitative discussion of the
long--range proximity effect on the LP oscillations in a
superconducting ring lying on a thin F (see
Fig.~\ref{Fig:1-Ring}(a)). We assume that $R_1 \gg \xi_f \gg w$.
In the absence of equal-spin triplet superconducting correlations,
proximity--induced superconductivity in F occupies the region $w_s
\approx w + 2 \xi_f \ll R_1$ under the S ring
(Fig.~\ref{Fig:1-Ring}(b)). As a result, the period of the LP
oscillations $\Delta H^{(s)}$ is modified due to the small
increase of the ring width $w_s \gtrsim w$ where superconductivity
coexists \cite{Groff-Parks-pr68}, i.e.,
$$
    \Delta H^{(s)} \sim \frac{\Phi_0}{\pi R_1^2 \left[ 1 + ( w_s /R_1)^2 \right]}
 %       \simeq \frac{\Phi_0}{\pi R_1^2} \left[ 1 - ( w_s / R_1)^2 \right]
        \,,
$$
and a slow modulation of the amplitude of the LP oscillations
takes place \cite{Samokhvalov-PRB09v}. Here $\Phi_0 = \pi \hbar c
/ |e|$ is the magnetic flux quantum.

Long-range equal--spin triplet pairs exist over a distance of the
order the thermal length $\xi_n = \sqrt{\hbar D_f / 2 \pi T_{cs}}
\gg \xi_f$, and proximity--induced superconductivity in F occupies
the region $w_t \approx \xi_n \gtrsim R_1$
(Fig.~\ref{Fig:1-Ring}(c)). Here, $T_{cs}$ is the transition
temperature of the superconductor in the absence of a proximity
effect, i.e. $T_c$ of S without F and an applied magnetic field.
In this case the spin--triplet pairs dominate a considerable part
of the S/F structure, as shown in Fig.~\ref{Fig:1-Ring}{c}. It has
recently been demonstrated that the spin--triplet odd--frequency
superconducting correlations emerging in layered S/F systems favor
the formation of the in-plane FFLO phase with the gap potential
modulated along the S/F interface
\cite{Mironov-PRL12_Meissner-effect,Mironov-PRL18_LOFF}. A
hallmark of in–plane FFLO instability is a vanishing of the London
magnetic field penetration depth $\lambda(\mathbf{r})$ averaged
over the structure volume. In multiply connected S/F structures
the in–plane FFLO instability is expected to provoke a
modification of $T_c(H)$ which correspond to the switching between
modes with the different vorticity $L$. This effect should be
especially important if $R_1 \gg \xi_f$, because of the mechanism
of vortex states switching (caused by the oscillatory behavior
spin-singlet pair wave function in a ferromagnet
\cite{Samokhvalov-PRB07v}) in this case is suppressed. The
experimental observation of the unusual behavior of the
Little-Parks oscillations predicted here would provide direct
evidence of spin-polarized triplet Cooper pairs.
%generation and, as a consequence, could be considered as a step in understanding
%the physics of the long-range proximity effect in these systems.

The paper is organized as follows. In Sec. II we describe the
model of a multiply connected S/F system and briefly discuss basic
equations. In Sec. III we proceed with the analytic calculations
of $T_c$ for different vortex states.
% on the basis of the linearized Usadel equations.
In Sec. IV we study the proximity induced switching between
different vortex states in the absence of an applied magnetic
field. We show that transitions between states with different
vorticity are accompanied by jumps of effective magnetic field
penetration depth. Moreover we analyse the influence of the
long-range spin-triplet correlations on the realisation of the
states with higher vorticities. The interplay between the
oscillations of $T_c$ due to the LP effect and oscillations due to
the $h$ is analyzed in Sec. V. We summarize our results in Sec.
VI.

\section{Model and basic equations}\label{ModelSection}

%%
%%%%%%%%%%%%%%%%%%%%%%%%%%%%%%%%%%%%%%%%%%%%%
\begin{figure}[t!]
\includegraphics[width=0.4\textwidth]{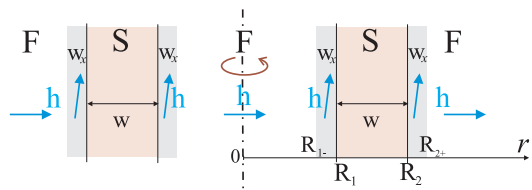}
\caption{(Color online) Cross section of the model cylindrical
structure under consideration. %
The blue arrows show the direction of the exchange field
$\mathbf{h}$
in the ferromagnet (F). %
} \label{Fig:2-FStruct}
\end{figure}
%%%%%%%%%%%%%%%%%%%%%%%%%%%%%%%%%%%%%
%%
The planar F/S system under consideration (Fig.~\ref{Fig:1-Ring}(a))
was shown to be qualitatively similar to a hollow superconducting
cylinder surrounded by a ferromagnetic metal
\cite{Samokhvalov-PRB07v}. Hereafter, we consider a model S/F
system consisting of a thin-walled hollow S cylinder embedded in a
ferromagnetic metal. The cross section of the model S/F structure
is shown in Fig.~\ref{Fig:2-FStruct}. The surrounding F metal has
a uniform exchange field $\mathbf{h}$ and spin--active S/F
interfaces. A good electrical contact between the F and S metals
is assumed to provide a strong proximity effect. The S and F
metals have diffusion coefficients $D_{s}$ and $D_{f}$,
respectively, and satisfy the dirty limit where $T_{cs} \tau /
\hbar \ll 1$ and $h \tau / \hbar \ll 1$. Here, $\tau = l / v_F$ is
the electron elastic scattering time with the electron mean free
path $l$. To observe a pronounced influence of the proximity
effect on $T_c$, the thickness of the S shell $w = R_2 - R_1$ must
be smaller than the superconducting coherence length $\xi_s =
\sqrt{\hbar D_s / 2 \pi T_{cs}}$. The exchange field $\mathbf{h}$
acting on the spin of the conduction electrons is assumed to have
step-like profile %%
\begin{equation}\label{eq:5}
    \mathbf{h}(r)=\left\{
            \begin{array}{cc}
                h\, \check{z}\,, &
                            \begin{array}{c}
                                r \le R_1 - w_x  \\
                                r \ge R_2 + w_x
                            \end{array} \,, \\
                h\, \check{z} + h_x(r)\, \check{x}\,, &
                            \begin{array}{c}
                                R_1 - w_x \le r \le R_1 \\
                                R_2 \le r \le R_2 + w_x
                            \end{array} \,, \\
                0\,, & R_1 < r \le R_2\,,          \end{array}
                \right.
\end{equation}
where $\check{x}$, $\check{z}$ are the axes for spin quantization.
We neglect reduction of the magnetization in the ferromagnet and
magnetization leakage into the superconductor due to the magnetic
proximity effect
\cite{Krivoruchko-PRB02,Bergeret-PRB04,Garifullin-JETPL11}. The
spin-active interface between S and F is described by the rotation
of the exchange field $\mathbf{h}$ in thin cylindrical layers $R_1
- w_x \le r \le R_1$ and $R_2 \le r \le R_2 + w_x$ near F/S
interfaces provided $w_x \ll \xi_f, R_1$ but the product $\int
dr\, h_x(r) / w_x h$ is of the order unity
\cite{Houzet-Buzdin-PRB07}.

Our calculations of the second-order superconducting phase
transition temperature $T_c$ are based on solutions of the
linearized Usadel equations \cite{Usadel-PRL70}.  At $T \sim T_c$,
the normal Green function $g$ coincides with the value in the
normal state ($g \simeq - i\, \mathrm{sgn}\, \omega_n$), and the
linearized Usadel equations for anomalous quasiclassical Green's
function \cite{Eschrig-RPP15,Mineev-Samokhin-Intro,Tokuyasu-PRB85}
\begin{equation}\label{eq:0}
    \hat{f}_{s,f} = f_{s,f}^s  + \mathbf{f}_{s,f}^t \hat\sigma\,, \qquad
    \mathbf{f}_{s,f}^t = f_{s,f}^z\,\mathbf{z}_0 + f_{s,f}^x\, \mathbf{x}_0
\end{equation}
take the form (see reviews
\cite{Buzdin-RMP05,Bergeret-RMP05,Champel-Eschrig-PRB05a} for
details):
\begin{eqnarray}\label{eq:1}
    &&- \frac{\hbar D_{s,f}}{2}\,\left( \nabla - \frac{2 i e }{\hbar c} \mathbf{A} \right)^2 f_{s,f}^s + \vert\, \omega_n\, \vert \,
    f_{s,f}^s  \label{eq:1a} \\
    &&\qquad\qquad\qquad +\, i\,{\rm sgn}\,\omega_n\, ( \mathbf{h}\, \mathbf{f}_{s,f}^t ) = \Delta\,,
    \nonumber \\
   &&- \frac{\hbar D_{s,f}}{2}\,\left( \nabla - \frac{2 i e }{\hbar c} \mathbf{A} \right)^2 \mathbf{f}_{s,f}^t + \vert\, \omega_n\, \vert \,
    \mathbf{f}_{s,f}^t  \label{eq:1b} \\
    &&\qquad\qquad\qquad +\, i\,{\rm sgn}\,\omega_n\, \mathbf{h} \, f_{s,f}^s  = 0\,.   \nonumber
\end{eqnarray}
Here, where $f_{s,f}^s(\mathbf{r},\omega_n)$
($\mathbf{f}_{s,f}^t(\mathbf{r},\omega_n)$) is the singlet
(triplet) part of the Green's function (\ref{eq:0}) in the
superconductor and ferromagnet, $\omega_n = (2 n + 1) \pi T_c$ is
the Matsubara frequency at the temperature $T_c$, $\Delta =
\Delta_s$ is the singlet pairing potential inside superconductor
($\Delta=0$ in ferromagnet), and $\mathbf{A}$ is the vector
potential of the external magnetic field $\mathbf{H} =
\mathrm{rot}\mathbf{A} = H \mathbf{z}_0$. We have neglected the
effect of the ferromagnet magnetization $\mathbf{M}$ because of
the additional magnetic flux enclosed by the S shell $\Phi_M
\approx 4\pi^2 R_1^2 M$  is assumed to be small in comparison with
the magnetic flux quantum $\Phi_0 = \pi \hbar c / |e|$ for typical
parameters of S/F hybrids \cite{Samokhvalov-PRB09v}. The Usadel
equations (\ref{eq:1a}),(\ref{eq:1b}) must be supplemented with
the Kupriyanov-Lukichev boundary conditions
\cite{Kuprianov-Lukichev-JETP88} for all components of the Green
function (\ref{eq:0}) at the S/F interfaces $r=R_1$ and $r=R_1$:
\begin{eqnarray}\label{eq:2}
    \sigma_s\, \partial_{r} \hat{f}_s
    = \sigma_f\, \partial_{r} \hat{f}_f\,, \qquad
    \hat{f}_s = \hat{f}_f + \gamma_b \xi_n\, \partial_{r}
    \hat{f}_f \,.
\end{eqnarray}
Here $\sigma_{f}$ and $\sigma_{s}$ are the normal-state
conductivities of the F and S metals, $\gamma_b$ is related to the
transparency of the S/F interface and is determined by the boundary
resistance per unit area $R_b$: $\gamma_b \xi_s = R_b \sigma_f$.

The $T_c$ is determined by the self-consistency equation for the
singlet gap function $\Delta_s$:
\begin{equation} \label{eq:4}
    \Delta_s(\mathbf{r})\,\ln \frac{T_c}{T_{cs}}
    + \pi T_c \sum_{\omega_n} \left(\frac{\Delta_s(\mathbf{r})}{ \omega_n }
        - f_s^s(\mathbf{r},\omega_n)\right) = 0\,.
\end{equation}
For simplicity of $T_c$ calculations, we assume $h \gg T_{cs}$ and
neglect proximity suppression caused by a finite S/F interface
resistance, i.e., we take $\gamma_b = 0$ in (\ref{eq:2}). In this
regime, $\hat{f}_s = \hat{f}_f$ at the S/F interface.

Choosing the cylindrical coordinate system $(r, \theta, z)$ and
the gauge $\mathbf{A}=(0,\, A_\theta,\, 0)$, $A_\theta = r H / 2$
we consider homogeneous along $z$ solutions of the equations
(\ref{eq:1a}),(\ref{eq:1b}), characterized by a certain angular
momentum $L$, referred further as vorticity
\begin{eqnarray}\label{eq:6}
    \Delta_s(\mathbf{r}) = \Delta_L(r)\, \mathrm{e}^{i L \theta}\,,
    \:
    f^{s,z,x}(\mathbf{r},\omega_n) = f_L^{s,z,x}(r,\omega_n)\,
        \mathrm{e}^{i L \theta }\,.
\end{eqnarray}
The vorticity parameter $L$ in (\ref{eq:6}) coincides with the
angular momentum of the Cooper pair wave function.

According to Eqs.~(\ref{eq:1a}),(\ref{eq:1b}),(\ref{eq:4}) there
is a symmetry $f^s(r,-\omega_n)=f^s(r,\omega_n)$ and
$f^{z,x}(r,-\omega_n)=-f^{z,x}(r,\omega_n)$, so that we can treat
only positive $\omega_n$ values. The Usadel equations
(\ref{eq:1a}) and (\ref{eq:1b}) can be written in the form
\begin{eqnarray}
    &&-\frac{\hbar D}{2} \hat{Q}_L f_L^s\,
        + \omega_n\, f_L^s + \,\imath\, h\, f_L^z + \,\imath\, h_x\, f_L^x = \Delta_L, \label{eq:8a} \\
    &&-\frac{\hbar D}{2} \hat{Q}_L f_L^x
        +\, \omega_n\, f_L^x + \,\imath\, h_x\, f_L^s = 0\,,            \label{eq:8b} \\ %
    &&-\frac{\hbar D}{2} \hat{Q}_L f_L^z
        +\, \omega_n\, f_L^z + \,\imath\, h\, f_L^s  = 0\,,             \label{eq:8c}  \\
    &&\quad\qquad\hat{Q}_L = \frac{1}{r}\,\frac{d}{d r}\left( r\,\frac{d}{d r}\,\right)
        - \left(\frac{L - \phi_r}{r}\right)^2                           \nonumber
\end{eqnarray}
where $D = D_{s,f}$ is the diffusion coefficient in S and F,
respectively, $\phi_r = \pi r^2 H / \Phi_0$ is a dimensionless
flux of the external magnetic field $H$ threading the circle of
certain radius $r$. As follows from
Eqs.~(\ref{eq:8a}),(\ref{eq:8b}),(\ref{eq:8c}), if the gap
potential $\Delta_L$ is real, the components $f_L^s$ of the
anomalous Green’s function are also real, while the components
$f_L^x$ and $f_L^z$ are imaginary. Then it is convenient to
introduce the complex function
\cite{Mironov-PRB15HM} % S. Mironov, A. Buzdin, Phys. Rev. B \textbf{92}, 184506 (2015)
\begin{equation} \label{eq:10}
    F(r) = f_L^s + i\,f_L^z
\end{equation}
and the real function
\begin{equation} \label{eq:11}
    P(r) = -i\, f_L^x\,,
\end{equation}
so that
\begin{equation} \label{eq:12}
    f_L^s = \mathrm{Re} [ F ] \,, \quad
    f_L^z = i\, \mathrm{Im}[ F ]\,, \quad \mathrm{and} \quad
    f_L^x = i\, P\,.
\end{equation}
These functions satisfy the equations %%
\begin{eqnarray}
    &&-\frac{\hbar D}{2} \hat{Q}_L F
        + \left( \omega_n\, + \imath\, h \right)\, F  - h_x\, P  = \Delta_L\,, \label{eq:13a} \\
    &&-\frac{\hbar D}{2} \hat{Q}_L P
        +\, \omega_n\, P + h_x\,\mathrm{Re} [ F ] = 0\,.          \label{eq:13b}
\end{eqnarray}
The boundary conditions for $F$ and $P$ follow directly from the
conditions (\ref{eq:2}) for $\hat{f}$ components.

\section{$T_c$ of vortex states}\label{CritTempSection}

We proceed with $T_c$ calculations for different vortex states
(\ref{eq:6}).

\subsection{Solution inside the F core: ($r \le R_1$)}                        %%%%%%%%%%%%%%%%%%%%

The solution of Eqs.~(\ref{eq:13a}),(\ref{eq:13b}) in the F
cylinder $r \le R_1 - w_x$ with uniform exchange field $\mathbf{h}
= h\, \mathbf{z}_0$ can be expressed via the confluent
hypergeometric function of the first kind (Kummer's function)
$K(a,b,z)$ \cite{Abramowitz-Handbook}
\begin{eqnarray}
    &&F_{f1}(r)= C_1\,\mathrm{e}^{-\phi_r/2} \phi_r^{\vert\,L\,\vert/2}\,
               K\left(\,a_{Ln},\,b_L,\,\phi_r\,\right)\,,        \label{eq:14a} \\
    &&P_{f1}(r)= C_2\,\mathrm{e}^{-\phi_r/2} \phi_r^{\vert\,L\,\vert/2}\,
               K\left(\,a_{Ln}^r,\,b_L,\,\phi_r\,\right)\,,     \label{eq:14b}
\end{eqnarray}
where
\begin{eqnarray}
    &&a_{Ln} = \frac{\vert L \vert - L + 1}{2} +
        \frac{(\omega_n + i\, h)\, R_1^2}{2 \hbar D_f\, \phi_1} \,,          \nonumber \\
    &&a_{Ln}^r = \mathrm{Re}[a_{Ln}]\,, \qquad
        b_L = \vert L \vert +1                                               \nonumber
\end{eqnarray}
and $\phi_1 = \pi R_1^2 H / \Phi_0$ is a dimensionless flux of the
external magnetic field $\mathbf{H}$ threading the circle of
the radius $R_1$.%
To proceed further with tractable formulas, we take into account
that the spin-active layer near the interface $r = R_1$ is thin
($w_x \ll \xi_f,\,R_1$). After averaging
Eqs.~(\ref{eq:14a}),(\ref{eq:14b}) over the thickness of thin $w_x
\rightarrow 0$ layer one can receive the following relations
between the values of the functions $F_{f1}$, $P_{f1}$ and their
derivatives $\partial_r F_{f1}$, $\partial_r P_{f1}$ at the S/F
interface $r = R_1$: %
\begin{eqnarray}
    &&\frac{d F_{f1}}{d r} = \frac{\kappa_{Ln}}{R_1}\, F_{f1}
           - \frac{2 \delta}{\xi_f}\, P_{f1}\,,                     \label{eq:17a} \\
    &&\frac{d P_{f1}}{d r} = \frac{\mu_{Ln}}{R_1}\, P_{f1}
           + \frac{2 \delta}{\xi_f}\, \mathrm{Re}[F_{f1}]\,,        \label{eq:17b}
\end{eqnarray}
where $\delta$, characterizing the spin-activity of
the S/F interface, is determined by %
\begin{equation}\label{delta}
    \delta = \frac{\xi_f}{\hbar D_f} \int\limits_{R_1-w_x}^{R_1} h_x(r) dr \,,
\end{equation}
and
\begin{eqnarray}
    \kappa_{Ln} &\equiv& \kappa_{Ln}(\phi_1) = \vert L \vert - \phi_1                     \label{eq:18a} \\
        &+& 2 \phi_1 \frac{a_{Ln}\, K\left(\,a_{Ln}+1,\,b_L+1,\,\phi_1\,\right)}
                                     {b_L\, K\left(\,a_{Ln},\,b_L,\,\phi_1\,\right)}\,,    \nonumber \\
    \mu_{Ln} &\equiv& \mu_{Ln}(\phi_1) = \vert L \vert - \phi_1               \label{eq:18b} \\
        &+& 2 \phi_1 \frac{a_{Ln}^r\, K\left(\,a_{Ln}^r+1,\,b_L+1,\,\phi_1\,\right)}
                                     {b_L\,K\left(\,a_{Ln}^r,\,b_L,\,\phi_1\,\right)}\,.  \nonumber
\end{eqnarray}

\subsection{Solution in outer ferromagnet: ($r \ge R_2$)}      %%%%%%%%%%%%%%%%%%%%

The solution of Eqs.~(\ref{eq:13a}),(\ref{eq:13b}) in ferromagnet
with a cylindrical cavity of radius $r = R_2$ can be expressed via
the confluent hypergeometric function of the second kind
$U(a,b,z)$ \cite{Abramowitz-Handbook}
\begin{eqnarray}
    &&F_{f2}(r)= \tilde{C}_1\,\mathrm{e}^{-\phi_r/2} \phi_r^{\vert\,L\,\vert/2}\,
               U\left(\,a_{Ln},\,b_L,\,\phi_r\,\right)\,,               \label{eq:19a} \\
    &&P_{f2}(r)= \tilde{C}_2\,\mathrm{e}^{-\phi_r/2} \phi_r^{\vert\,L\,\vert/2}\,
               U\left(\,a_{Ln}^r,\,b_L,\,\phi_r\,\right)\,.             \label{eq:19b}
\end{eqnarray}
As before the spin-active layer near the S/F interfaces ($r =
R_2$) is assumed to be thin and described by $\delta$
(\ref{delta}). The relations between the values of the functions
$F_{f2}$, $P_{f2}$ and their derivatives $\partial_r F_{f2}$,
$\partial_r P_{f2}$ at the S/F interfaces $r = R_2$ are as
follows:
\begin{eqnarray}
    &&\frac{d F_{f2}}{d r} = \frac{\tilde{\kappa}_{Ln}}{R_2}\, F_{f2}
           + \frac{2 \delta}{\xi_f}\, P_{f2}\,,                     \label{eq:20a} \\
    &&\frac{d P_{f2}}{d r} = \frac{\tilde{\mu}_{Ln}}{R_2}\, P_{f2}
           - \frac{2 \delta}{\xi_f}\, \mathrm{Re}[F_{f2}]\,,        \label{eq:20b}
\end{eqnarray}
where
\begin{eqnarray}
    \tilde{\kappa}_{Ln} &\equiv& \tilde{\kappa}_{Ln}(\phi_2) = \vert L \vert - \phi_2    \label{eq:23a} \\
     &-& 2 \phi_2 \frac{a_{Ln}\, U\left(\,a_{Ln}+1,\,b_L+1,\,\phi_2\,\right)}
                            {U\left(\,a_{Ln},\,b_L,\,\phi_2\,\right)}\,,               \nonumber  \\
    \tilde{\mu}_{Ln} &\equiv& \tilde{\mu}_{Ln}(\phi_2) = \vert L \vert - \phi_2   \label{eq:23b} \\
     &-& 2 \phi_2 \frac{a_{Ln}^r\, U\left(\,a_{Ln}^r+1,\,b_L+1,\,\phi_2\,\right)}
                              {U\left(\,a_{Ln}^r,\,b_L,\,\phi_2\,\right)}\,.          \nonumber
\end{eqnarray}
and $\phi_2 =  \pi R_2^2 H / \Phi_0$ is the flux of the external
magnetic field enclosed in the cavity of radius $R_2$ in units of
$\Phi_0$.

\subsection{Solution in thin superconducting shell: ($R_1 \le r \le R_2$)}      %%%%%%%%%%%%%%%%%%%%

In absence of an exchange field, equations (\ref{eq:13a}) and
(\ref{eq:13b}) in the superconducting region $R_1 \le r \le R_2$
take the form
\begin{eqnarray}
    & &-\frac{\hbar D_s}{2} \hat{Q}_L F_s
        + \omega_n\, F_s  = \Delta_L \,,    \label{eq:24a} \\
    & &-\frac{\hbar D_s}{2} \hat{Q}_L P_s
        +\, \omega_n\, P_s  = 0\,.          \label{eq:24b}
\end{eqnarray}
Assuming that variations of $F_s(r)$, $P_s(r)$ and $\Delta_L(r)$
in the thin S shell are small for $w \lesssim \xi_s, R_1$, we can
average Eqs.~(\ref{eq:24a}) and (\ref{eq:24b}) over the thickness,
using the Kupriyanov-Lukichev boundary conditions (\ref{eq:2}) and
relations (\ref{eq:17a}),(\ref{eq:17b})
(\ref{eq:20a}),(\ref{eq:20b}) to integrate the terms $\partial_r (
r\,\partial_r F_s )$ and $\partial_r ( r\,\partial_r P_s )$.
Finally, we obtain the following equations
\begin{eqnarray}
    &&\left( \bar{\Omega}_n + \nu_{L n} \right)\, F_s -
        \epsilon\, P_s = \bar{\Delta}_L \,,                         \label{eq:25a} \\
\nonumber \\
    &&\left( \bar{\Omega}_n  + \eta_{L n} \right)\, P_s +
        \epsilon\, Re [ F_s ] = 0 \,,                               \label{eq:25b}
\end{eqnarray}
where $\bar{\Omega}_n = \Omega_n / T_{cs}$, $\bar{\Delta}_L =
\Delta_L / T_{cs}$,
$$
    \Omega_n = \omega_n + \frac{D_s}{2} \frac{\left(\,L -
        \phi_1\,\right)^2}{R_1^2}\,, \:
      \beta = \frac{\pi\, \xi_s^2\, \sigma_f / \sigma_s}{w\, R_1}\,,
$$
$$
    \nu_{Ln} = \nu_{Ln}^r + i\,\nu_{Ln}^i
    = \beta \left(\kappa_{Ln} - \tilde{\kappa}_{Ln} \right) \,,
$$
$$
    \eta_{Ln} = \beta \left( \mu_{Ln} - \tilde{\mu}_{Ln} \right) \,,
\quad
    \epsilon = 2 \beta \delta \left( R_1 + R_2 \right) / \xi_f^2 \,.
$$
Solution of the algebraic system (\ref{eq:25a}),(\ref{eq:25b})
determines the amplitudes $\bar{f}_L^s$, $\bar{f}_L^z$,
$\bar{f}_L^x$ of the anomalous Green’s functions (\ref{eq:2}) in
superconductor for the orbital mode $L$:
%
%\begin{widetext}
\begin{eqnarray}
    &&\bar{f}_L^s = \frac{ \bar{\Delta}_L \left(\bar{\Omega}_n + \nu_{Ln}^r \right)}
    {\ds \vert \bar{\Omega}_n + \nu_{Ln} \vert^2
        + \epsilon^2 \left( \bar{\Omega}_n + \nu_{Ln}^r \right)
        / \left( \bar{\Omega}_n + \eta_{Ln} \right) } \,, \qquad                      \label{eq:32a} \\
    &&\bar{f}_L^z = \frac{-i\, \nu_{Ln}^i}{\bar{\Omega}_n + \nu_{Ln}^r} \bar{f}_L^s\,, \qquad
      \bar{f}_L^x = \frac{- i\, \epsilon}
            {\bar{\Omega}_n + \eta_{Ln}} \bar{f}_L^s   \,.                         \label{eq:32b}
\end{eqnarray}
%\end{widetext}
%
Substituting solution (\ref{eq:32a}) into Eq.~(\ref{eq:4}) one
obtains a self--consistency equation for the critical temperature
$T_L$ of the state with a vorticity $L$:
%
%%
%\begin{equation} \label{eq:4}
%    \ln \frac{T_L}{T_{cs}}
%    + \pi T_L \sum_{\omega_n} \left(\frac{1}{ \omega_n }
%        - \frac{1}{\ds \Omega_n + \beta\bar{\kappa}_{L1}
%        + \beta^2 \left( \frac{\ds \bar{\kappa}_{L2}^2}{\Omega_n + \beta \bar{\kappa}_{L1}}
%        + \frac{4\, (\delta_1 + \delta_2)^2 R_f^2}{\xi_f^4 (\Omega_n + \beta \bar{\kappa}_{Ln})}
%        \right)} \right) = 0\,.
%\end{equation}
%%
%
\begin{equation} \label{eq:33}
    \ln \frac{T_L}{T_{cs}}
    + \pi T_L \sum_{\omega_n} \left(\frac{1}{ \omega_n }
        - \frac{\bar{f}_L^s}{\Delta_L} \right) = 0\,.
\end{equation}
As usual, the critical temperature $T_c$ of a superconductivity
nucleation in the S/F hybrid is determined by the maximal value
$T_L$:
\begin{equation} \label{eq:34}
    T_c = \underset{L}{\rm max}\{T_L\}\,.
\end{equation}

\section{Proximity induced vortex states}\label{ProximitySection}

%
%%%%%%%%%%%%%%%%%%%%%%%%%%%%%%%%%%%%%%%%%%%%%
\begin{figure*}%[t!]
\includegraphics[width=0.45\textwidth]{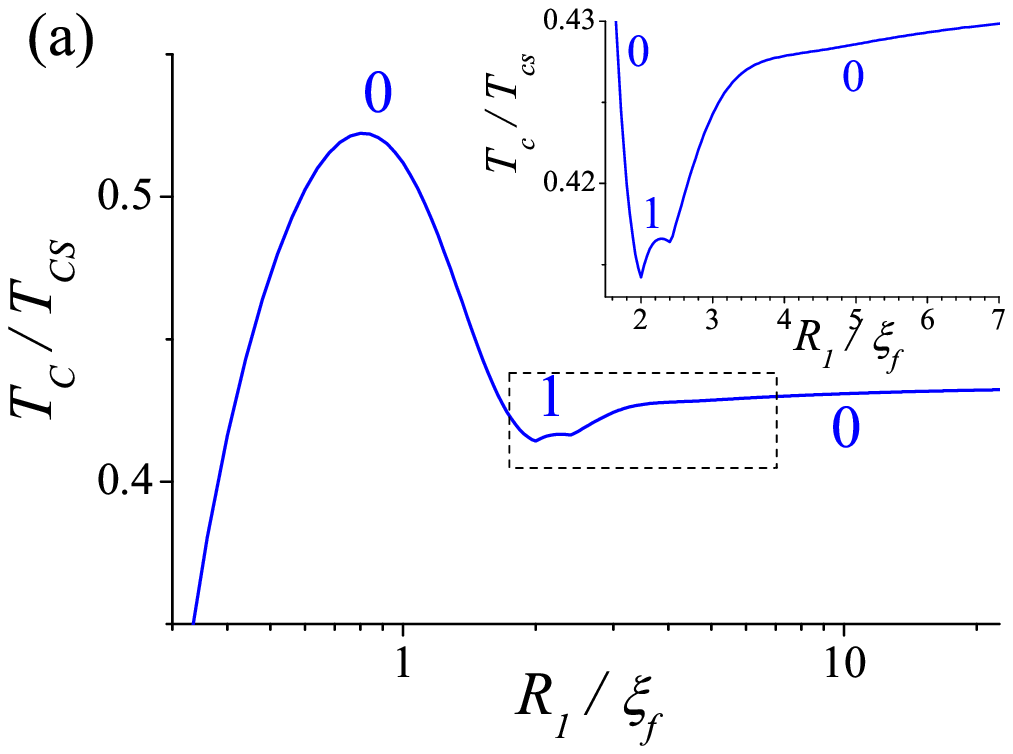}
\includegraphics[width=0.45\textwidth]{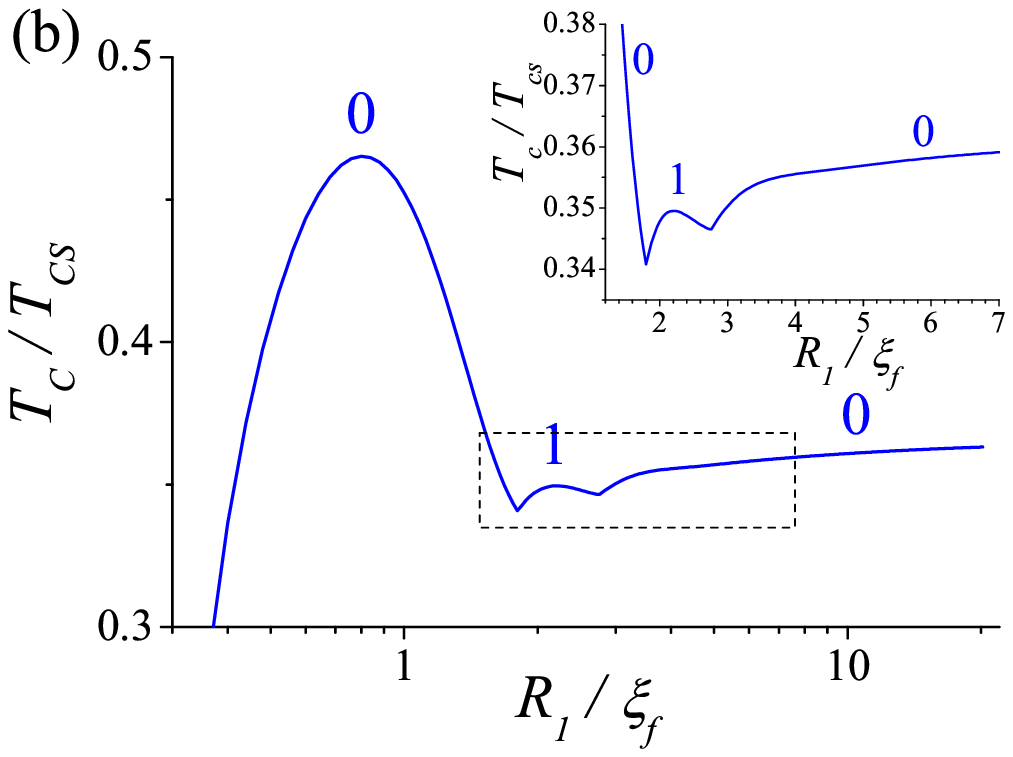}
\includegraphics[width=0.45\textwidth]{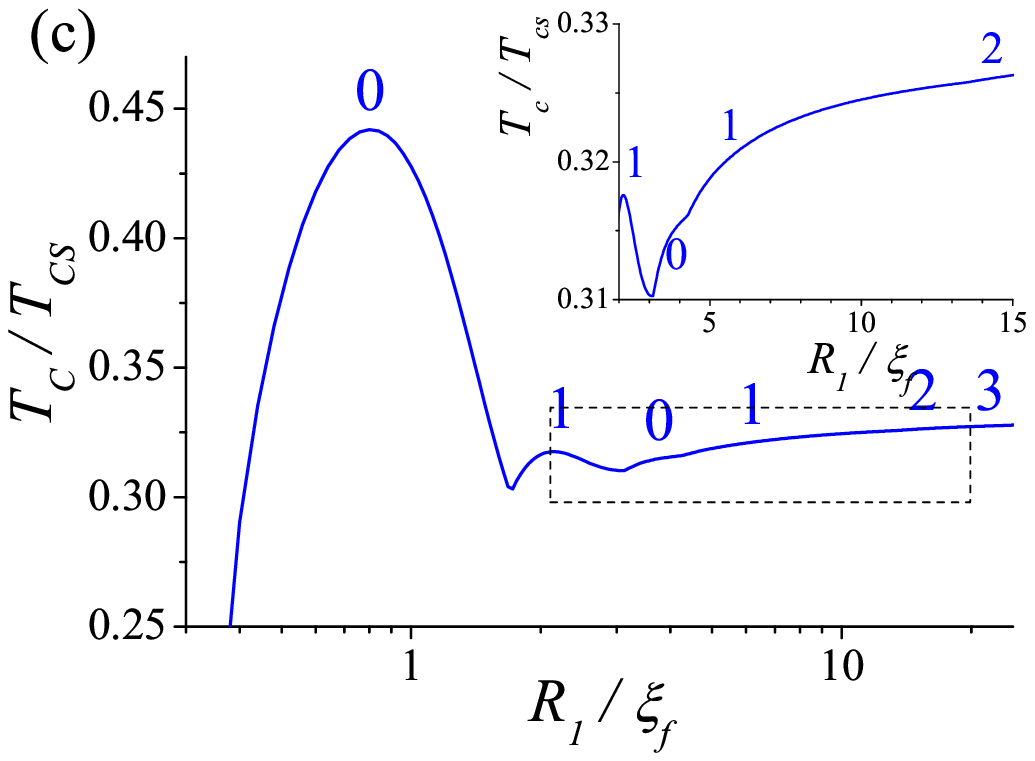}
\includegraphics[width=0.45\textwidth]{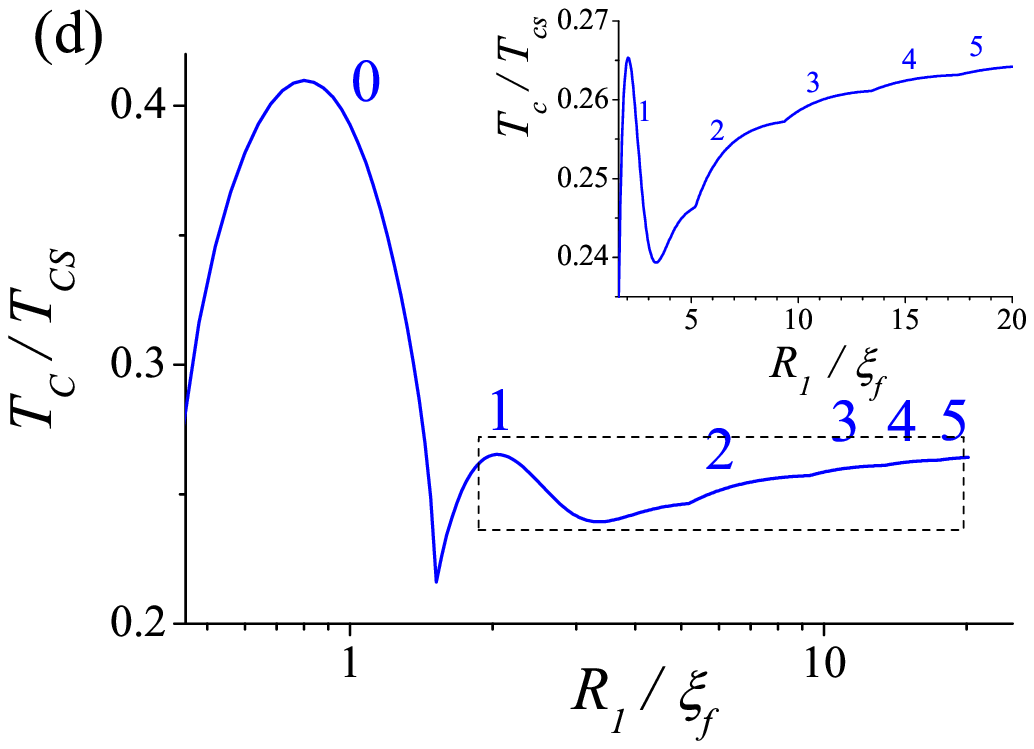}
\caption{(Color online) Typical dependences of $T_c$ on the
internal radius $R_1$ of the S ring in zero applied field ($H =
0$) for different values of $\delta=\delta_1+\delta_2$
characterizing the spin-activity of the S/F interface: ($a$)
$\delta=0$; ($b$) $\delta=0.25$; ($c$) $\delta=0.3$; ($d$)
$\delta=0.35$. The numbers near the curves denote the
corresponding vorticity $L$. The inserts show the zoom in  the
part of the curves in the rectangles. Here we choose $w =0.6\,
\xi_s$; $\xi_s / \xi_f = 0.1$; $\xi_n / \xi_f = 5$; $\sigma_f /
\sigma_s = 0.5$.} \label{Fig:3}
\end{figure*}
%%%%%%%%%%%%%%%%%%%%%%%%%%%%%%%%%%%%%
%%
We start our analysis for the case of zero external magnetic field
focusing on the effect of the S/F interface spin-activity and
long--range triplet generation on the phase transition line
$T_c(R_f)$. The solution of Eqs.~(\ref{eq:13a}),(\ref{eq:13b}) for
$H = 0$ in the ferromagnet can be expressed via the modified
Bessel functions $I_L(\zeta)$ and $K_L(\zeta)$ of the of the order
$L$:
\begin{eqnarray}
&&\begin{array}{cc}
     \begin{array}{c}
        F_{f1}(r) = \sum\limits_{\omega_n} \bar{F}_L(\omega_n) \frac{\ds I_L(k_n r)}{\ds I_L(k_n R_1)}\,,  \\
        P_{f1}(r) = \sum\limits_{\omega_n} \bar{P}_L(\omega_n) \frac{\ds I_L(q_n r)}{\ds I_L(q_n R_1)}\,,  \\
     \end{array}
        r \le R_1-w_x,\:
  \end{array}                                   \label{eq:35a} \\
&&\begin{array}{cc}
     \begin{array}{c}
        F_{f2}(r) = \sum\limits_{\omega_n} \bar{F}_L(\omega_n) \frac{\ds K_L(k_n r)}{\ds K_L(k_n R_2)}\,, \\
        P_{f1}(r) = \sum\limits_{\omega_n} \bar{P}_L(\omega_n) \frac{\ds K_L(q_n r)}{\ds K_L(q_n R_2)}\,, \\
     \end{array}
        r \ge R_2+w_x,\:
  \end{array}                                   \label{eq:35b}
\end{eqnarray}
where $\bar{F}_L = \bar{f}_L^s + \bar{f}_L^z$, $\bar{P}_L = -i
\bar{f}_L^x$, $q_n = \xi_n^{-1}\,\sqrt{T_c (2 n + 1) / T_{cs}}$,
$k_n^2 = q_n^2 + 2 i / \xi_f^2$. Taking into account the solutions
(\ref{eq:35a}),(\ref{eq:35b}) we obtain the following simplified
expressions for the parameters $\kappa_{L n}$ (\ref{eq:18a}),
$\mu_{Ln}$ (\ref{eq:18b}), $\tilde{\kappa}_{L n}$ (\ref{eq:23a}),
$\tilde{\mu}_{Ln}$ (\ref{eq:23b}):
\begin{eqnarray}
    &&\kappa_{L n} = \vert L \vert +
        \frac{(k_n R_1)\, I_{L+1}(k_n R_1)}{I_{L}(k_n R_1)}\,,        \nonumber  \\
    &&\mu_{L n} = \vert L \vert +
        \frac{(q_n R_1)\, I_{L+1}(q_n R_1)}{I_{L}(q_n R_1)}\,,  \nonumber  \\
    &&\tilde{\kappa}_{L n} = \vert L \vert -
        \frac{(k_n R_2)\, K_{L+1}(k_n R_2)}{K_{L}(k_n R_2)}\,,        \nonumber  \\
    &&\tilde{\mu}_{Ln} = \vert L \vert -
        \frac{(q_n R_2)\, K_{L+1}(q_n R_2)}{K_{L}(q_n R_2)}\,,  \nonumber
\end{eqnarray}
which have to be used in expressions to calculate the critical
temperature $T_L$ of the proximity induced state with a vorticity
$L$. Note, that the states with angular momenta $\pm L$ are
degenerated for $H = 0$ and have the same critical temperature
$T_{-L}=T_L$.

In Fig.~\ref{Fig:3} we show example dependencies of $T_c$ versus
$R_1$ for different values of $\delta$ characterizing the
spin-activity of the S/F interface. We see that generation of long
triplet correlations in F metal produces an additional depairing
effect which results in a overall decrease in $T_c$ for arbitrary
$L$. At the same time, the destructive action of long--range
proximity effect reduces the region of $R_1$ where the vortex free
state $L = 0$ dominates, and provokes the appearance of the states
with nonzero orbital momentum  $L \ne 0$. Note that for strong
enough spin-flip scattering at the S/F interface, the $T_c(R_1)$
phase boundary exhibits quasiperiodic oscillations as function of
the ring radius in the region $R_1 \gg \xi_f$ (see
Fig.~\ref{Fig:3}(c),(d)). In this case the influence of
damped-oscillatory behavior of the singlet $f_L^s$ and the short
triplet $f_L^{z}$ components of superconducting correlations is
weak and cannot provoke switching between the superconducting
states with different vorticities.

\begin{figure*}[t!]
\includegraphics[width=0.45\textwidth]{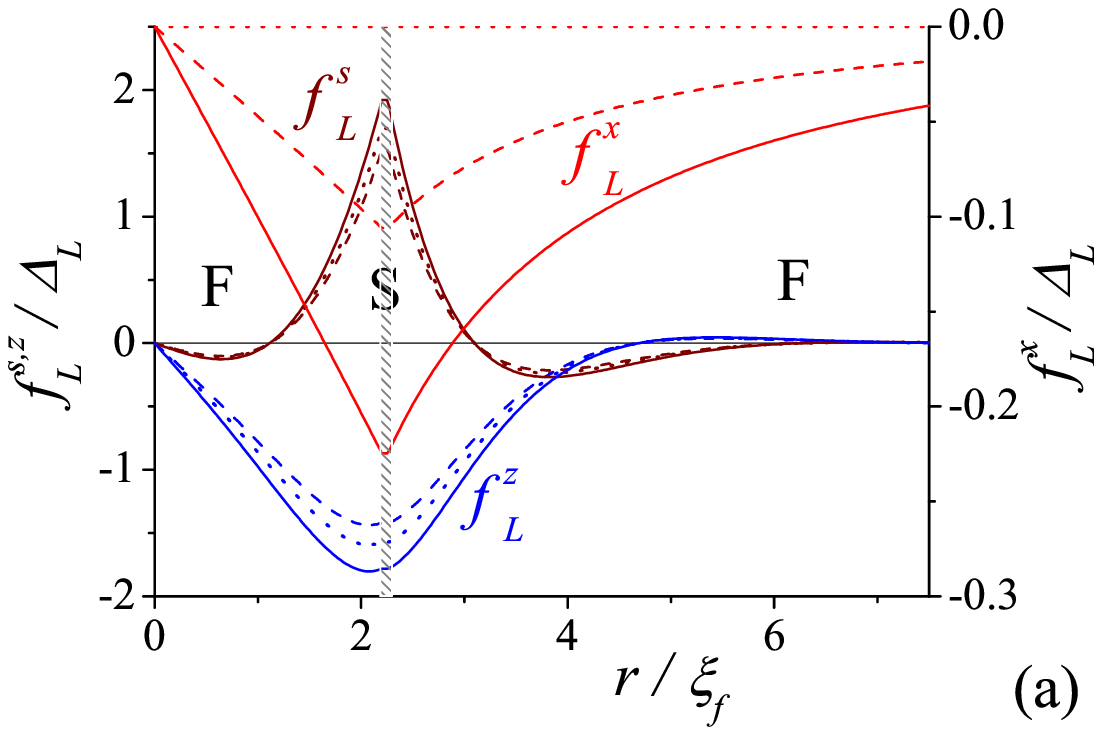}
\includegraphics[width=0.45\textwidth]{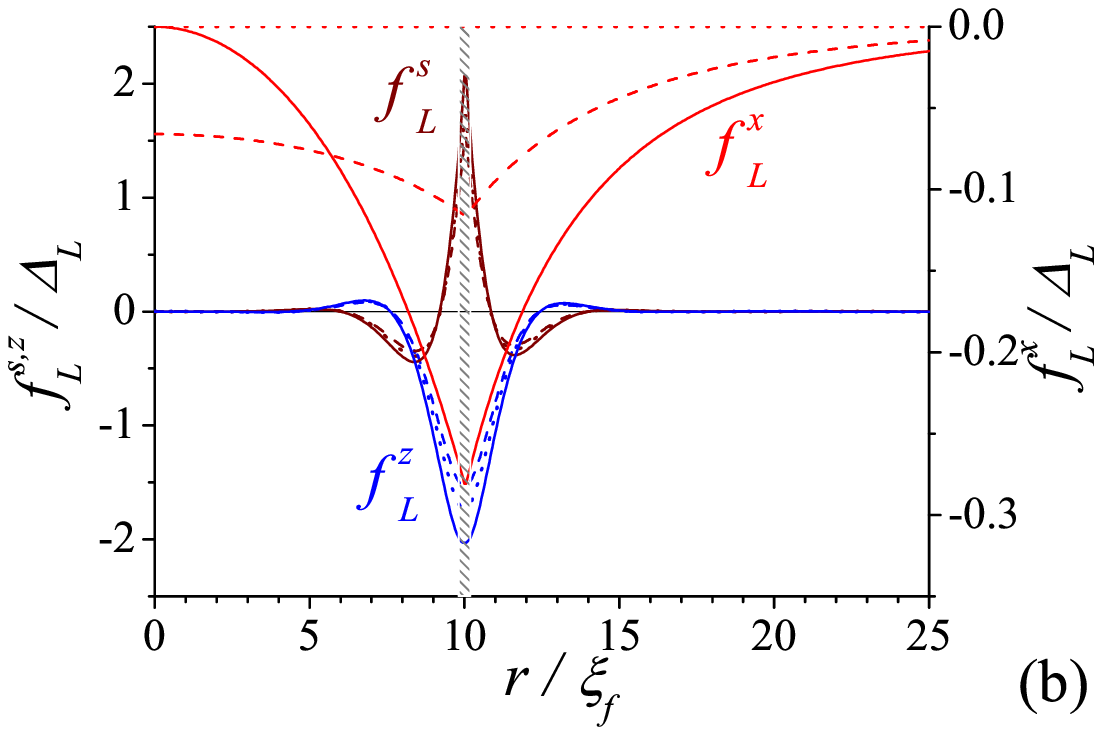}
\caption{(Color online) Spatial dependence of the components
$f_L^{s,z,x}$ (vorticity $L=1$) of the Green function $\hat{f}$
(\ref{eq:0}) in zero applied field ($H = 0$) for different values
of the the ring radius: ($a$) $R_1=2.2\,\xi_f$; ($b$)
$R_1=10.0\,\xi_f$ and for different values of the parameters
$\delta$ characterizing a spin-activity of the S/F interface:
$\delta=0$ (dotted line); $\delta=0.25$ (dashed line);
$\delta=0.35$ (solid line). %
The other parameters are the same as in Fig.~\ref{Fig:3}.}
\label{Fig:4}
\end{figure*}

In Fig.~\ref{Fig:4} we have plotted the spatial dependence of
$f_L^s$ and $f_L^{z,x}$ components of the quasiclassical Green's
function $\hat{f}$ (\ref{eq:0}) for two different radia of the
ring. One can see that $f_L^s$ and short--range triplet $f_L^z$
decay and oscillate on the scale $\xi_f$ while the long--range
triplet component $f_L^x$ decays to zero slowly. As a result, the
triplet component $f_L^x$ dominates over a considerable part of
the S/F structure. Note that recently it was demonstrated that the
spin-triplet superconducting correlations emerging in S/F system
favors the formation of the in-plane FFLO phase with the gap
potential modulated along the S/F interface
\cite{Mironov-PRL12_Meissner-effect}. In our case this FFLO-like
phase is revealed by the formation of the high vorticity states at
large $R_1$ radius - see Fig.~\ref{Fig:3}(c),(d). A hallmark of
in–plane FFLO instability is vanishing of the London magnetic
field penetration depth $\lambda(\mathbf{r})$ averaged over the
structure volume.
%
%%%%%%%%%%%%%%%%%%%%%%%%%%%%%%%%%%%%%%%%%%%%%
%%%%%%%%%%%%%%%%%%%%%%%%%%%%%%%%%%%%%
%%

\subsection{Effective magnetic field penetration depth}

To calculate the London penetration depth $\lambda(\mathbf{r})$
that appears in relation
\begin{equation}\label{eq-London}
    \mathbf{j} = \frac{c}{4 \pi \lambda^2}\,\left( \frac{\hbar c}{2 e} \nabla \phi -
    \mathbf{A} \right)\,,
\end{equation}
between screening current density $\mathbf{j}$ and vector
potential $\mathbf{A}$, we can use the expression for the $k$
projection of the supercurrent in terms of quasi-classical Usadel
functions $\hat{f}(\mathbf{r},\omega_n)$ defined by the
parametrization (\ref{eq:0}) \cite{Eschrig-RPP15}
\begin{equation}\label{eq-ScreenCurr}
    j_{s,f}^k = \pi T_c \frac{\sigma_{s,f}}{e} \sum_{\omega_n} \mathrm{Im}
        \left[ (f_{s,f}^s)^*\,\widetilde{\nabla}_k f_{s,f}^s
        - (\mathbf{f}_{s,f}^t)^*\,\widetilde{\nabla}_k \mathbf{f}_{s,f}^t
        \right]\,,
\end{equation}
and the local London penetration depth
\begin{equation}\label{eq:lambda0}
    \lambda(\mathbf{r})=\left\{
            \begin{array}{cc}
                \lambda_f(r)\,, & r < R_1\,, \quad r > R_2\,, \\
                 \lambda_s\,, & R_1 \le r \le R_2\,, \\
            \end{array}
                \right.
\end{equation}
depends on the radius $r$ and can be expressed via the amplitudes
$F(r)$ and $P(r)$ as follows:
\begin{equation}\label{eq:lambda}
    \lambda_{s,f}^{-2}(r) = \frac{8 \pi^2 T_L }{\hbar c^2}
        \sum_{\omega_n} \sigma_{s,f} \left[\, Re [ F_{s,f}^2(r) ] - P_{s,f}^2(r) \right]
        \,,
\end{equation}
where $F_s = F_L(\omega_n)$, $P_s = P_L(\omega_n)$, $F_f(r) =
F_{f1}(r)$ ( $F_{f2}(r)$ ) and $P_f(r) = P_{f1}(r)$ ( $P_{f2}(r)$
) for $r \le R_1$ ( $r \ge R_2$ ), respectively. If triplet
superconducting correlations dominate in a ferromagnetic region,
the corresponding value of $\lambda_f^{-2}(r)$ becomes negative,
and the local screening current (\ref{eq-ScreenCurr}) is
paramagnetic. At the same time, $\lambda_s^{-2}(r) > 0$ in the
thin S shell and $\lambda_f^{-2}(r)
> 0$ in ferromagnetic layers $\sim \xi_f$ near the S/F boundaries,
where the direction of the screening current corresponds to the
conventional diamagnetic Meissner effect. A superconducting state
remains stable in the whole as long as the London penetration
depth (\ref{eq:lambda}) averaged over the cross section of the S/F
structure
\begin{equation}
    \Lambda^{-2} = \frac{2 \pi}{S} \int\limits_0^{R_\infty}
        \frac{r dr}{\lambda^{2}(r)} \,, \qquad S = \pi R_\infty^2
\end{equation}
is negative \cite{Mironov-PRL12_Meissner-effect}. The external
radius of the structure $R_\infty$ is assumed to be large enough
($R_\infty \gg \xi_n$) to neglect the effect of the external
boundary on the solutions (\ref{eq:19a}),(\ref{eq:19b}).
Substituting the solutions of the linearized Usadel equations
(\ref{eq:32a}),(\ref{eq:32b}),(\ref{eq:35a}),(\ref{eq:35b}) into
(\ref{eq:lambda}), we obtain the following expression for the
effective magnetic penetration depth $\Lambda_L$ of the orbital
mode $L$ that the temperature $T$ is close to the critical
temperature $T_L$:
\begin{widetext}
\begin{eqnarray}
    \Lambda_L^{-2} &=& \Lambda_T^{-2} \sum_{\omega_n}
    \left\{ \mathrm{Re} \left[ \bar{F}_L^2(\omega_n) \left(\frac{w}{R_1}
        + \frac{\sigma_{f}}{\sigma_{s}}C_L(k_n) \right)  \right]
       - \bar{P}_L^2(\omega_n) \left(\frac{w}{R_1}
        + \frac{\sigma_{f}}{\sigma_{s}}C_L(q_n) \right) \right\} \,,    \label{eq:Lambda} \\
     \Lambda_T^{-2} &=& \frac{16 \pi^3 T \sigma_{s} R_1^2}{\hbar c^2\, S}\,, \quad
     C_L(q) = \frac{R_2^2}{R_1^2} \frac{K_{L-1}(q R_2)\, K_{L+1}(q R_2)}{2\,K_{L}^2(q R_2)}
        - \frac{I_{L-1}(q R_1)\, I_{L+1}(q R_1)}{2\,I_{L}^2(q R_1)} -
        \frac{w}{R_1} \,.     \nonumber
\end{eqnarray}
\end{widetext}

In the case of a second-order phase transition at $T = T_L$ from
the superconducting to the normal state, the order parameter
$\Delta_L$ disappears ($\Delta_L,\,\bar{F}_s,\,\bar{P}_s \to 0$),
and effective penetration depth (\ref{eq:Lambda}) diverges
($\Lambda_L \to \infty$). At $T < T_L$, the structure of the
superconducting order parameter corresponding to the free energy
minimum may differ from (\ref{eq:6}), and the equilibrium value of
$\Delta_L$ in the S/F cylinder must be determined with the help of
complete nonlinear Usadel equations without using the linear
approximation (\ref{eq:1a}),(\ref{eq:1b}). Therefore, expression
(\ref{eq:Lambda}) establishes only the relation between
superconducting order parameter $\Delta_L$ and the effective
magnetic field penetration depth $\Lambda_L$ in the S/F structure
for $L$ orbital mode at a temperature $T$ close to superconducting
transition temperature $T_L$.
%
%%%%%%%%%%%%%%%%%%%%%%%%%%%%%%%%%%%%%%%%%%%%%
\begin{figure*}%[t!]
\includegraphics[width=0.45\textwidth]{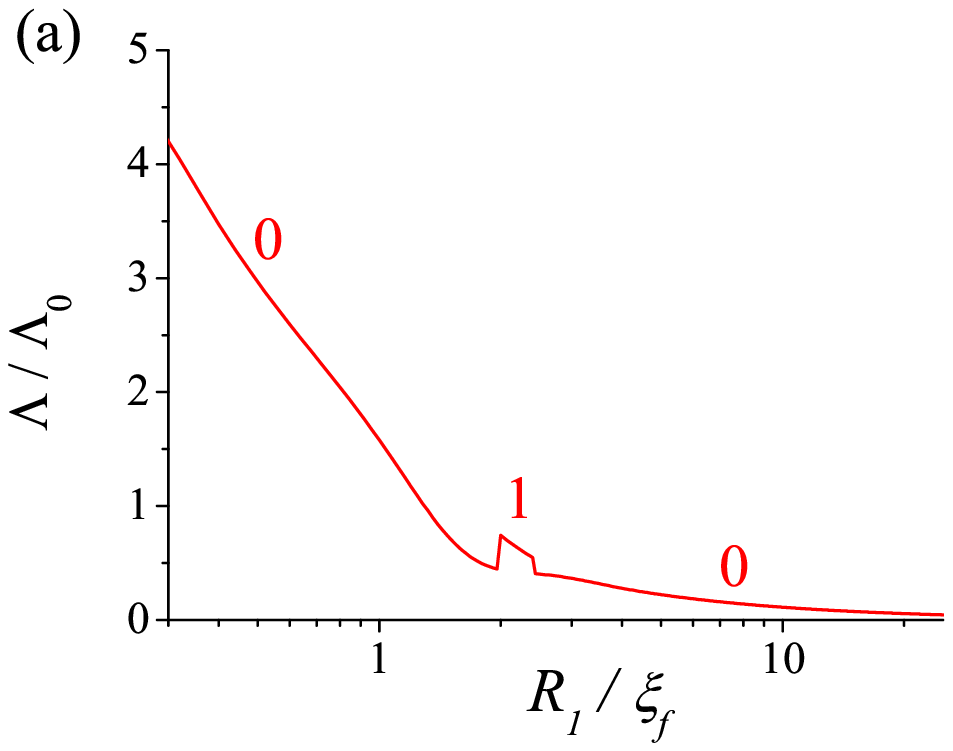}
\includegraphics[width=0.45\textwidth]{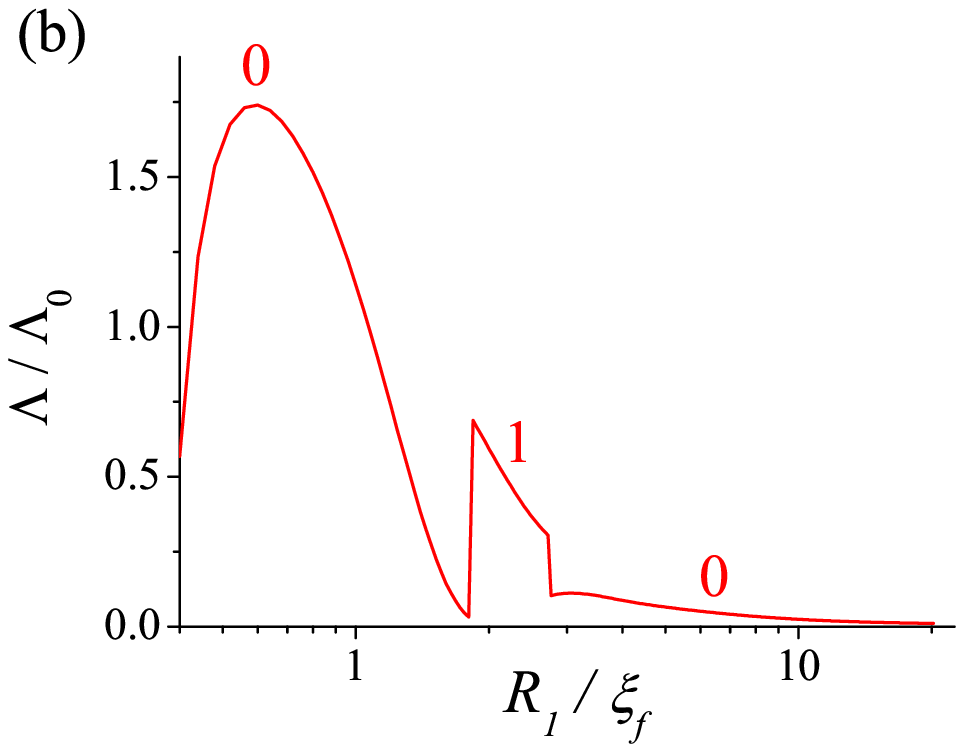}
\includegraphics[width=0.45\textwidth]{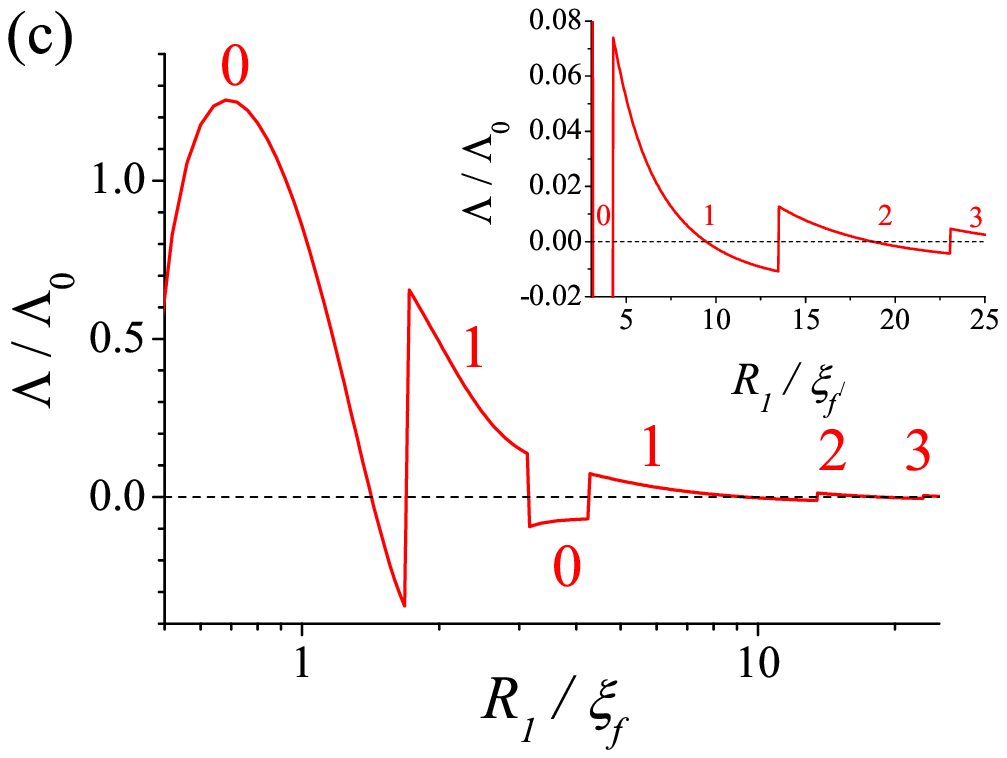}
\includegraphics[width=0.45\textwidth]{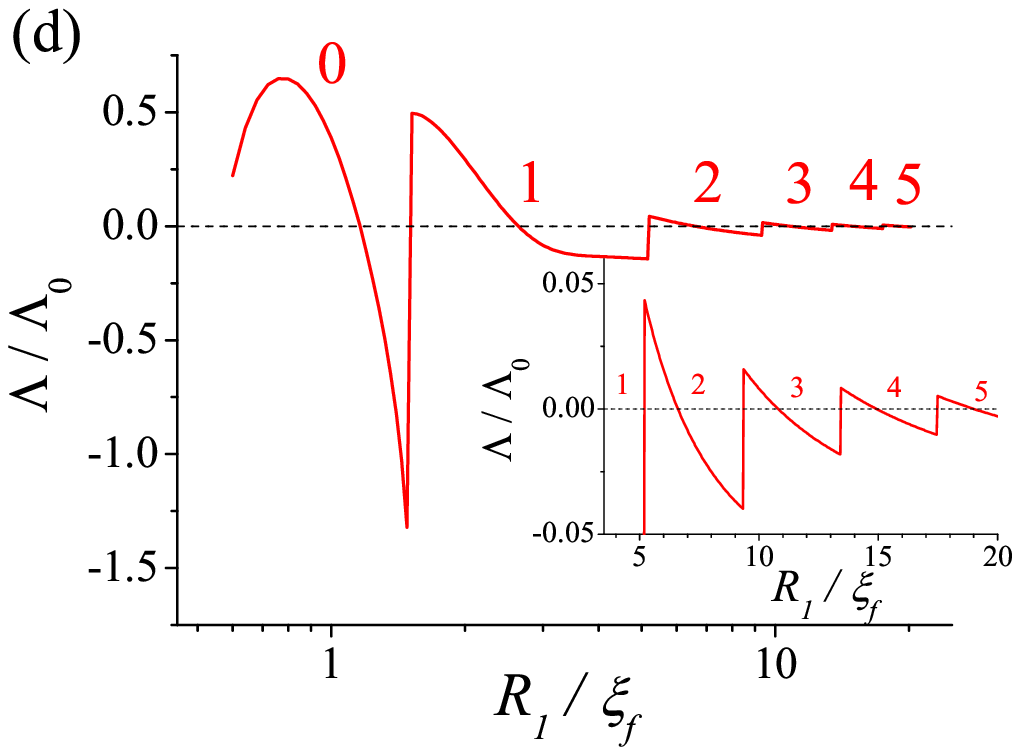}
\caption{(Color online) Typical dependences of the effective
penetration depth  $\Lambda$ the internal radius $R_1$ of the S
ring in zero applied field ($H = 0$) for different values of the
parameters $\delta$ characterizing a spin-activity of the S/F
interface: %
($a$) $\delta=0$; ($b$) $\delta=0.25$; ($c$) $\delta=0.3$; ($d$)
$\delta=0.35$. %
The numbers near the curves denote the corresponding values of
vorticity $L$. The inserts show the zoom in  the part of the
curves in the rectangles. The other parameters are the same as in
Fig.~\ref{Fig:3}.} \label{Fig:5}
\end{figure*}
%%%%%%%%%%%%%%%%%%%%%%%%%%%%%%%%%%%%%
%

Figure \ref{Fig:5} shows the typical dependence of the effective
penetration depth $\Lambda$ on $R_1$ for different values of
$\delta$. One can easily see that switching between different
orbital modes $L \rightleftarrows L + 1$ (see Fig.~\ref{Fig:3})
are accompanied by jumps in $\Lambda/\Lambda_T$ due to qualitative
changes in the radial structure and symmetry of the pair
wave-functions and the singlet/triplet correlations distributions
in the S/F structure \cite{Samokhvalov-JETP17}. Such abrupt
changes of screening properties of S/F cylinder due to orbital
modes switchings look similar to the $0\,-\,\pi$ transition in S/F/S
trilayer
\cite{Pompeo-Samokhvalov-PRB14,Samokhvalov-PRB15,Lemberger-PRB16}.
Negative values of $\Lambda$ correspond to the total paramagnetic
response of the hybrid structure. Here we neglect the possibility
of LOFF modulation along the S/F cylinder axis assuming that the
homogeneous in $z$ state exists in the planar hybrid structure
under consideration (see Fig.~\ref{Fig:1-Ring}). The effect of an
in-axis LOFF instability on screening properties of a thin--walled
superconducting cylinder filled with a ferromagnetic metal was
studied in \cite{Samokhvalov-JETP17}. In contrast to switching of
orbital modes $L \rightleftarrows L + 1$, the appearance of LOFF
modulation along the $z$ axis is not accompanied by abrupt changes
in the screening properties of the S/F cylinder and results in the
restoration of the diamagnetic response of the hybrid structure.
For layered S/F structures similar effect was first analyzed in
\cite{Mironov-PRL12_Meissner-effect}.

\section{Little-Parks oscillations}

%%%
%%%%%%%%%%%%%%%%%%%%%%%%%%%%%%%%%%%%%%%%%%%%%%
\begin{figure}%[t!]
\includegraphics[width=0.45\textwidth]{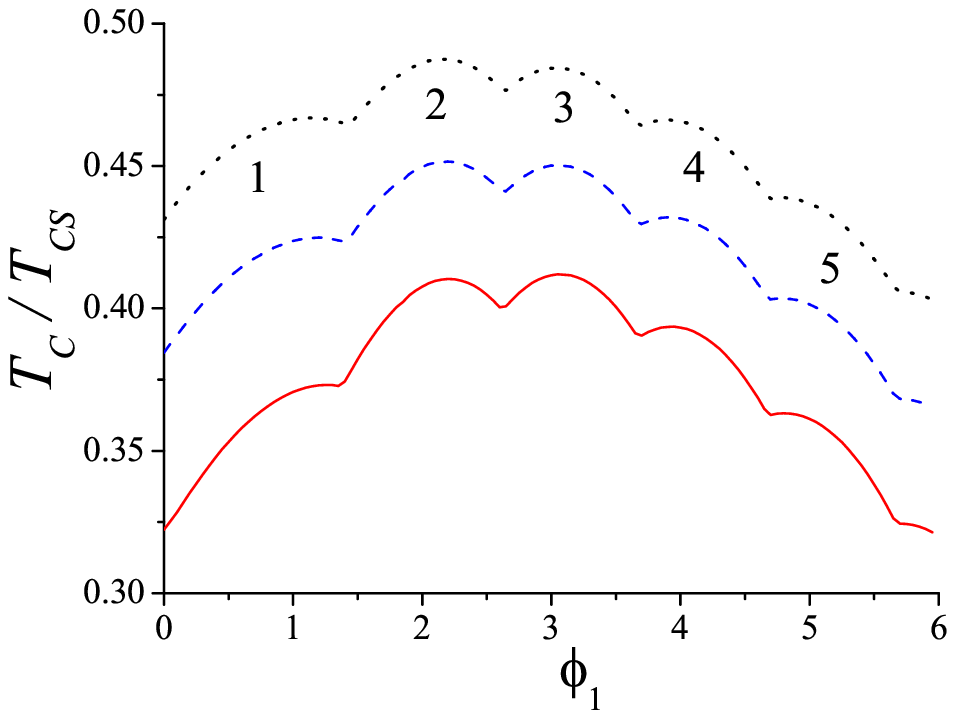}
\caption{(Color online) The dependences of the critical
temperature $T_c$ on the external magnetic field $H$ for different
values of the parameter $\delta$ characterizing a spin-activity of
the S/F interface: %
$\delta=0$ (dotted line); $\delta=0.25$ (dashed line);
$\delta=0.35$ (solid line). %
The radius of the ring $R_1 = 2.2\,\xi_f$. The magnetic field $H$
is measured in the units of the magnetic flux $\phi_1$ enclosed in
F core. The other parameters are the same as in Fig.~\ref{Fig:3}.
}\label{Fig:6}
\end{figure}
%%%%%%%%%%%%%%%%%%%%%%%%%%%%%%%%%%%%%%%%%%%%%%
%
%%%%%%%%%%%%%%%%%%%%%%%%%%%%%%%%%%%%%%%%%%%%
\begin{figure*}%[t!]
\includegraphics[width=0.45\textwidth]{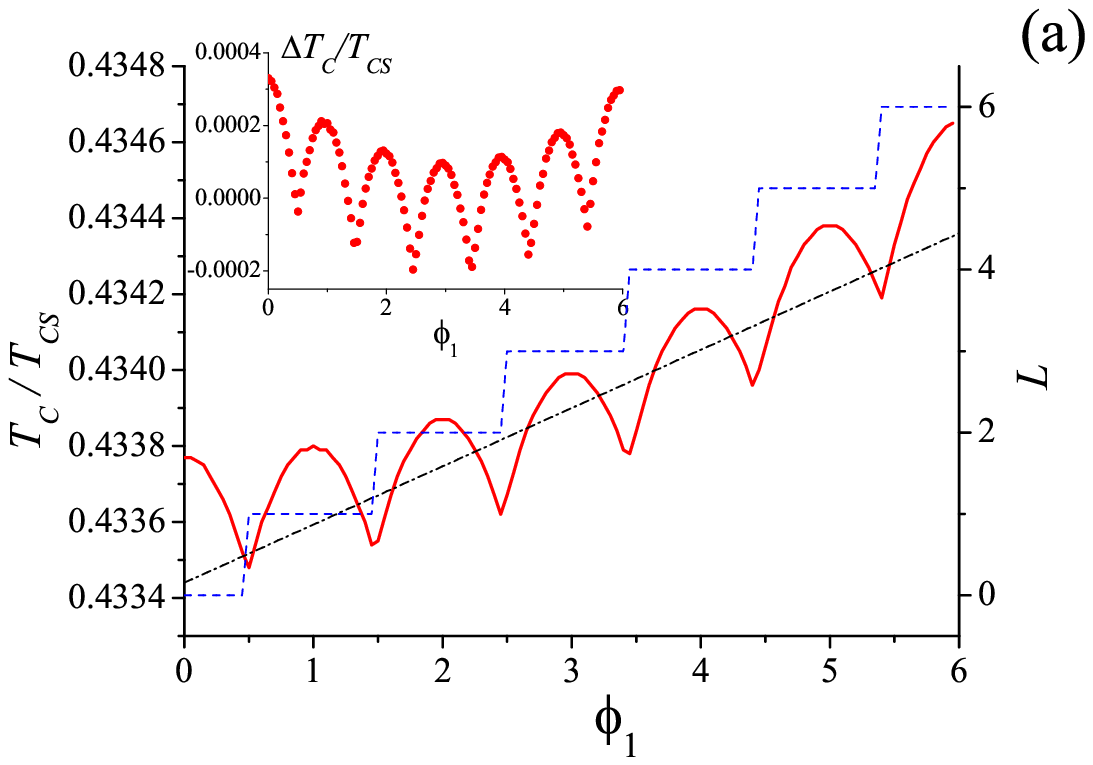}
\includegraphics[width=0.45\textwidth]{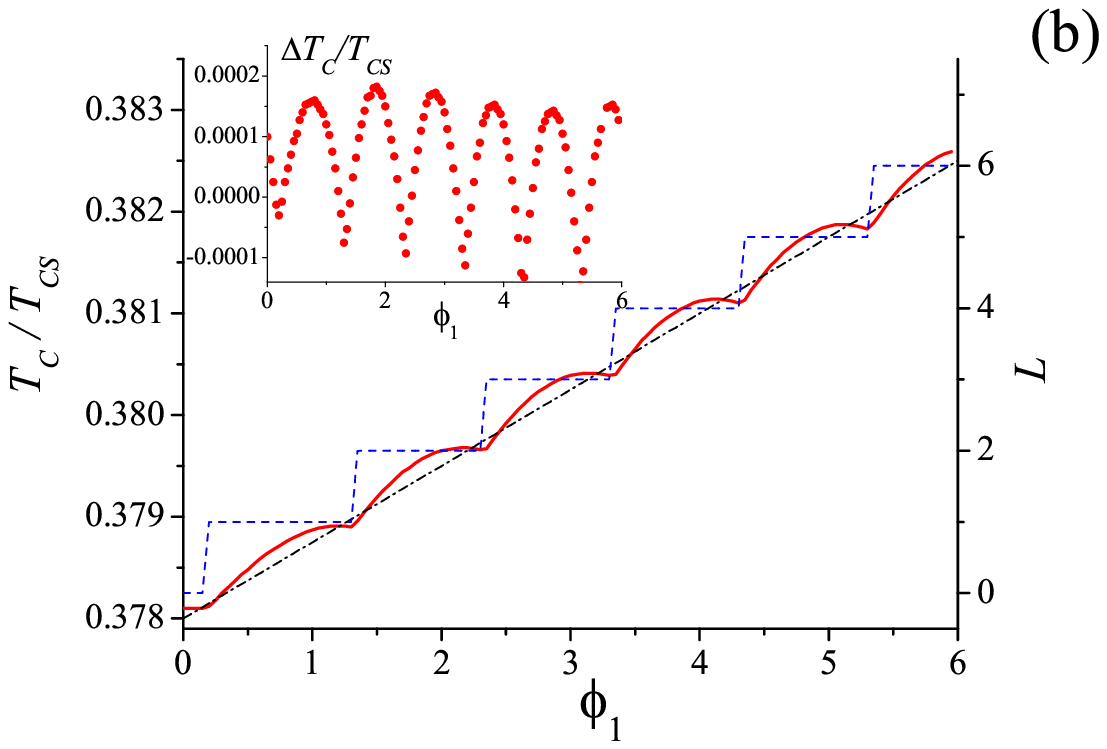}
\includegraphics[width=0.45\textwidth]{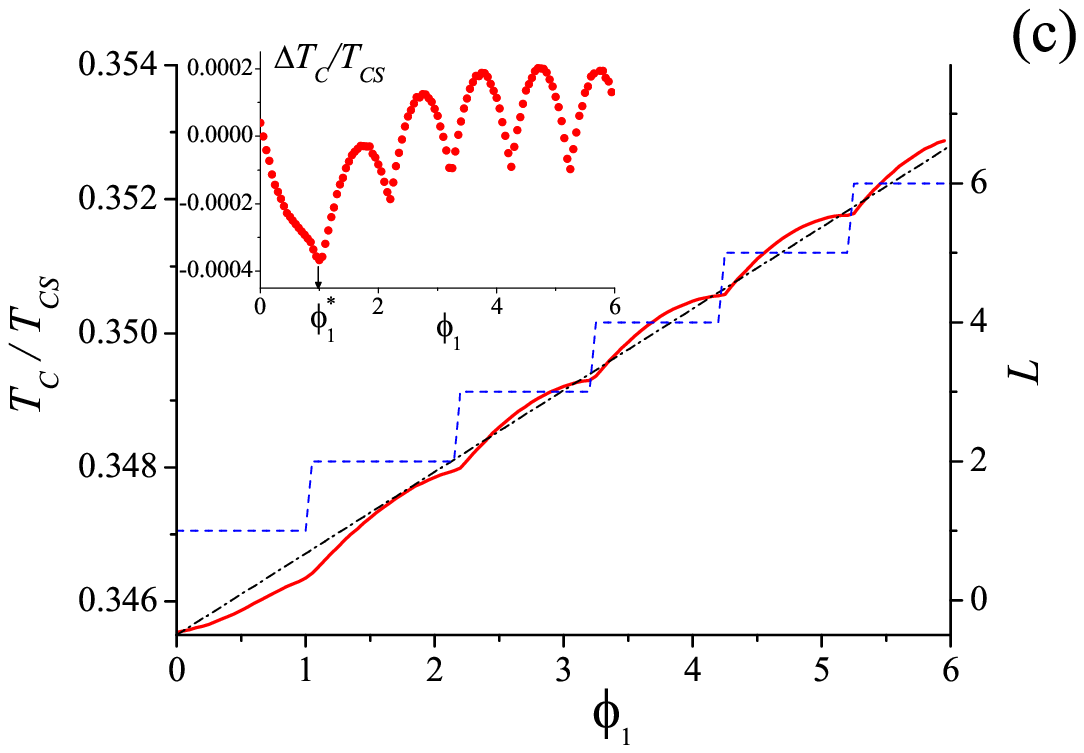}
\includegraphics[width=0.45\textwidth]{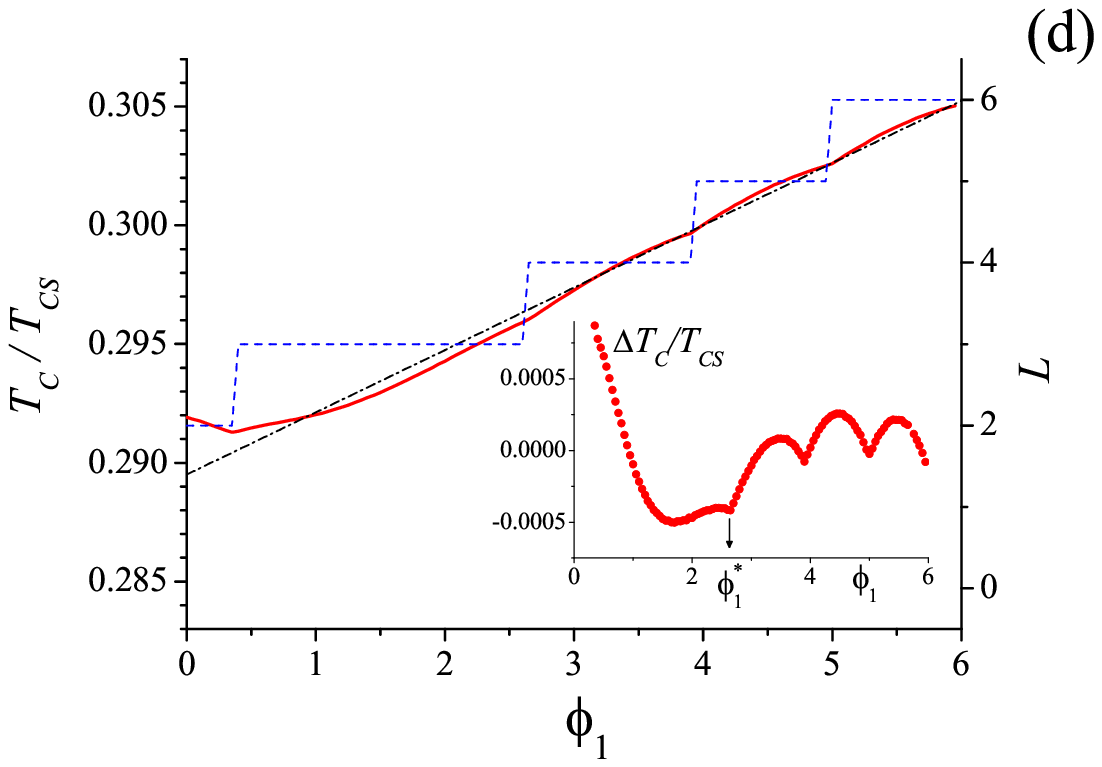}
\caption{(Color online) The dependences of the critical
temperature $T_c$ (solid line) and the vorticity $L$ (dashed line)
on the external magnetic field $H$ for different values of the
parameter $\delta$ characterizing the spin-activity of the S/F
interface: %
($a$) -- $\delta=0$; ($b$) -- $\delta=0.25$; ($c$) --
$\delta=0.3$; ($d$) -- $\delta=0.35$. %
The inserts show the dependence $\Delta T_c(\phi_1) = T_c(\phi_1)
- ( c_1 \phi_1 + c_0 )$ (the functions $c_1 \phi_1 + c_0$ are
shown by the dash-dotted line). The internal radius of the ring
$R_1 = 10\xi_f$. The magnetic field $H$ is measured in the units
of the magnetic flux $\phi_f$ enclosed in F core. The other
parameters are the same as in Fig.~\ref{Fig:3}. }\label{Fig:7}
\end{figure*}
%%%%%%%%%%%%%%%%%%%%%%%%%%%%%%%%%%%%
%%
Now we proceed with calculations of $T_c$ on external magnetic
field $\mathbf{H}$ applied along $z$ axis.
Figures~\ref{Fig:6},\ref{Fig:7} show example of dependency of
$T_c$ on the external magnetic flux $\phi_1 = \pi R_1^2 H /
\Phi_0$ for different values of $\delta$. The phase boundary
exhibits Little-Parks oscillations due to transitions between the
states with different angular momenta $L \leftrightarrows L \pm 1$
of the superconducting order parameter. For $R_1 \sim \xi_f$, the
$T_c(H)$ phase boundary exhibits quasiperiodic oscillations versus
$H$ (Figure~\ref{Fig:6}). One can observe a shift of the main
$T_c$ maximum toward nonzero $H$. The main maximum of $T_c$
corresponds to the minima of both the orbital and exchange effects
of superconductivity destruction. Since for chosen parameters of
the S/F structure the exchange part of the depairing effect is
minimal and the corresponding $T_c$ is maximal for a state with a
nonzero vorticity $L$, the orbital effect is cancelled at $\phi_1
\approx L \ne 0$. Thus, the main $T_c$ maximum shifts to a certain
nonzero $\phi_1$  value \cite{Samokhvalov-PRB09v}. Generation of
long triplet correlations in F metal produces an additional
depairing effect which results in a total decrease in the critical
temperature $T_c$ and a reduction of the interval of magnetic
field in which magnetism and superconductivity coexist.

Figure~\ref{Fig:7} shows the LP quasiperiodic oscillations of
$T_c$ versus $\phi_1$ for $R_1 \gg \xi_f$ and for different values
of $\delta$. One can observe a noticeable modification of the
$T_c(\phi_1)$ phase boundary when triplet superconducting
correlations dominate in ferromagnetic regions (see panels (c)
and (d) in Fig.~\ref{Fig:7}) -- the LP oscillations are
destroyed for small values of the magnetic flux. To manifest this
effect the inserts in Figure~\ref{Fig:7} show the dependence
$\Delta T_c(\phi_1) = T_c(\phi_1) - ( c_1\,\phi_1 + c_0 )$ where
the constants $c_{0,1}$ are chosen to compensate monotonic growth
of $T_c$ with increasing of the magnetic flux. One can see that
for $\delta=0.3$ and $\delta=0.35$ the curves $\Delta T_c(\phi_1)$
show the Little--Parks oscillations if the magnetic flux $\phi_1$
is large enough: $\phi_1 \ge \phi_1^*$, and the oscillations
are destroyed for smaller values magnetic flux $\phi_1$.%
The value of cutoff parameter $\phi_1^*$ grows with increasing the
spin-active constant $\delta$.

\section{Summary}\label{Summary}

We have analyzed switching between superconducting states with
different vorticities caused by the exchange field in multiply
connected S/F hybrids associated with generation of odd-frequency
spin-triplet correlations near S/F interface. As an example, we
have considered mesoscopic thin-walled superconducting cylindrical
shell embedded in ferromagnetic metal. A good electrical contact
and the spin-active S/F interface between the S and F metals are
assumed to assure a rather strong long ranged proximity effect. We
suggest a mechanism of switching between superconducting states
with different vorticities caused by prevalence of spin-triplet
pairs in a considerable part of the S/F structure. The spin-active
interface favors the emergence of the states with high vorticity
$L > 1$. The screening properties of the mesoscopic S/F structure
with a multiply connected geometry have been analyzed. The
presence of spin-triplet superconducting correlations results in
suppression of a diamagnetic response so that the effective
magnetic field penetration depth $\Lambda^{-2}$ can take a
negative value indicating a paramagnetic Meissner effect. The
observation of a paramagnetic response in the S/F setup of
Fig.~\ref{Fig:1-Ring}$(a)$ would provide clear evidence the
long-ranged odd--frequency triplet correlations. The behavior of
the Little--Parks oscillations of the critical temperature $T_c$
on an external magnetic flux $\phi_1$ threading the ring was
analyzed. The interplay between the orbital and exchange effects
results in breaking of the periodicity of $T_c(\phi_1)$ dependence
and a slow modulation of the amplitude of the quasiperiodic
oscillations. %
The Little--Parks oscillations are shown to be destroyed in the
region of small values of the magnetic flux threading the ring if
a spin-activity of S/F interface is strong enough so that the
spin-triplet pairs prevail in the multiply connected hybrid. %
%
%With an increase in spin-activity of S/F interface we observe
%destruction of the Little--Parks oscillations for small values of
%the magnetic flux $\phi_1$ threading the ring.

\acknowledgments

This work was supported by the Russian Science Foundation under
the Grant No.15-12-10020 ( A.V.S.), and the French ANR project
SUPERTRONICS and OPTOFLUXONICS (A.I.B.). A.V.S. appreciate
warm hospitality of the University Bordeaux, extended to him during
the visits when this work was done. J.W.A.R. acknowledges
funding from the EPSRC through an International Network Grant and
Programme Grant (No. EP/P026311/1 and No. EP/N017242/1). J.W.A.R.
also acknowledges funding from the Royal Society through a
University Research Fellowship and with A.I.B., funding from the
Leverhulme Trust and EU Network COST CA16218 (NANOCOHYBRI).

\end{document}